\documentclass{article}
\usepackage{latexsym}
\usepackage{amssymb,amsmath,amsfonts,mathtools}
\usepackage{commath}
\usepackage{subfigure}
\usepackage{enumerate}
\usepackage{slashed}
\usepackage{bbm}
\usepackage{epsfig}
\usepackage{wrapfig}
\usepackage{yfonts}
\usepackage{geometry}  
\usepackage{physics}
\usepackage{color}
\definecolor{hyperref}{RGB}{026,028,185}
\usepackage[bookmarks=true,colorlinks=true,linkcolor=hyperref,citecolor=hyperref,urlcolor=hyperref,bookmarksnumbered]{hyperref}
\usepackage{multirow}

\usepackage[sort,compress]{cite}

\newcommand{\grp}[1]{\mathrm{#1}}

\newcommand{\grSL}{\grp{SL}}
\newcommand{\grU}{\grp{U}}

\newcommand{\R}{\mathbb{R}}

\begin{document}
	\title{Hidden Conformal Symmetry in Higher Derivative Dynamics}
	
	\author{
		\large{
			 \href{mailto:vgmp@hi.is}{Valentina Giangreco M. Puletti}
			 and 
			 \href{mailto:vlmartin@hi.is}{Victoria L. Martin}}\\[5mm]
		{\normalsize University of Iceland, Science Institute}\\
		{\normalsize Dunhaga 3, 107 Reykjav{\'i}k, Iceland}
	}

	\date{}

	\maketitle

	\noindent\textbf{Abstract.}  The Kerr/CFT correspondence provides a holographic description of spinning black holes that exist in our universe, and the notion of hidden conformal symmetry allows for a formulation of this correspondence away from extremality.  In this work we study how hidden conformal symmetry is manifest when we consider dynamics beyond the Klein-Gordon equation, through studying the analytic structure of higher derivative equations of motion of a massless probe scalar field on a Kerr background using the monodromy method. Since such higher derivative dynamics appear in known examples of holographic AdS/logCFT correspondences, we investigate whether or not a Kerr/logCFT correspondence might be possible.

	\pagebreak 
	
	\tableofcontents
	\section{Introduction}
	Studying the nature of black holes has led to some of the deepest and most fruitful physical insights of the last century. The realization that black holes possess an entropy proportional to their surface area \cite{PhysRevD.7.2333, hawking1975particle} was the first step toward identifying the holographic nature of spacetime \cite{tHooft:1993dmi, Susskind:1994vu, Bousso:2002ju}: information stored in any spacetime volume is related to the surface area bounding that region. The most famous concrete realization of this principle in string theory is the Anti-de Sitter/Conformal Field Theory (AdS/CFT) correspondence \cite{Maldacena:1997re,Witten:1998qj}, which relates a theory of gravity in negatively-curved spacetime to a highly-symmetric quantum field theory in fewer spacetime dimensions. Holographic dualities like AdS/CFT are the workhorse of modern theoretical physics, harnessing black hole physics to study a surprisingly diverse plethora of phenomena, including strongly interacting condensed matter systems \cite{Hartnoll:2016apf}, the geometrization of entanglement \cite{ryu2006holographic,ryu2006aspects,maldacena2013cool}, Hawking evaporation \cite{Penington:2019npb,Almheiri:2019psf} and quantum chaos \cite{Shenker:2013pqa,Maldacena:2015waa}.
	
	Remarkably, holographic correspondences now exist for real, physical black holes that appear in our universe, which are maximally (or near-maximally) spinning Kerr black holes \cite{kerr1963gravitational}. The Kerr/CFT correspondence \cite{Guica:2008mu} relies on taking a near-horizon limit in the extremal Kerr metric, and showing that the isometries of the resulting metric form a copy of the conformal algebra, $\grSL(2,\mathbb R)$. A version of the Kerr/CFT correspondence was also found away from the extremal limit \cite{Castro:2010fd}. The authors of \cite{Castro:2010fd} found that the price of moving away from extremality is that it is necessary to consider symmetries of the near-horizon dynamics (the wave equation), rather than only symmetries of the near-horizon metric, to uncover an underlying conformal algebra. Such symmetries of the dynamics are referred to as \textit{hidden symmetries}, to distinguish them from explicit isometries of the metric. This notion of hidden conformal symmetry associated with black hole horizons has provided new insight into interesting and potentially observable aspects of black holes, such as black hole shadows \cite{hioki2008hidden}, tidal Love numbers \cite{Charalambous:2021kcz} and near-superradiant geodesics~\cite{Porfyriadis:2016gwb}. Furthermore, the presence of hidden symmetries in a gravitational system are known to be responsible for the separability of equations of motion on that background, as well as complete integrability of geodesics \cite{Frolov:2017kze}.
	
	A particularly useful method of examining near-horizon dynamics of a probe scalar field on a black hole background is to study the monodromy properties of solutions to the Klein-Gordon equation \cite{Castro:2013kea,Castro:2013lba,Aggarwal:2019iay,Chanson:2020hly}. The monodromy data encode information about black hole thermodynamics and the hidden conformal symmetry of \cite{Castro:2010fd}, and provide ample evidence for a two-dimensional CFT description of the thermal properties of black hole microstates. In particular, the fact that hidden conformal symmetry appears to be a feature of a large class of black holes \cite{Castro:2013kea,Sakti:2020jpo,Keeler:2021tqy} seems to indicate that a Cardy formula \cite{Cardy:1986ie} may be sensible to apply and able to reproduce the Bekenstein-Hawking entropy in scenarios beyond four- and five-dimensional black holes \cite{Strominger:1997eq,Castro:2010fd,Krishnan:2010pv}. 
	
	The goals of this work are twofold. First, since hidden conformal symmetry is only discernible through studying dynamics of a probe field, we begin this article by asking: how is hidden conformal symmetry manifest when we change the dynamics? That is, we would like to study an action such that the resulting equations of motion of our probe field are not the standard Klein-Gordon equation. In addition, we would like our chosen dynamics to potentially yield novel physical insight, while at the same time being easily comparable to the known Klein-Gordon case. To this end, we consider the higher derivative action of a massless scalar field $\Phi$:
	\begin{equation}\label{Introaction}
		S=-\int d^{4}x\sqrt{g}\left(\frac{1}{2}\Phi\left(\nabla^{\mu}\nabla_{\mu}\right)^n\Phi\right),
	\end{equation}
	with corresponding equation of motion 
	\begin{equation}\label{Introgenreq}
		\left(	\nabla^{\mu}\nabla_{\mu}\right)^n\Phi=0.
	\end{equation}
	When the integer $n=1$, we recover the familiar Klein-Gordon equation for a free scalar field. Higher order differential equations like this one arise in other physical settings, such as in the study of buoyant thermal convection \cite{littlefield1990frobenius}.
	
	Although the equation of motion \eqref{Introgenreq} is a very simple extension of the standard Klein-Gordon case, these higher derivative interactions already possess interesting physical attributes. Holographic duals of logarithmic conformal field theories (logCFTs) \cite{Gurarie:1993xq} are known to involve higher derivative equations of motion \cite{Hogervorst:2016itc}. LogCFTs are interesting in their own right, with applications in percolation \cite{cardy1999logarithmic} and quenched disorder \cite{cardy1999logarithmic,Caux:1995nm,Maassarani:1996jn,Caux:1998sm,Cardy:2013rqg}. Thus our second motivating question behind looking for hidden conformal symmetry in the dynamics \eqref{Introgenreq} is: can we construct a holographic logCFT correspondence in the spirit of Kerr/CFT (in which the conformal symmetry exists at the black hole horizon) as opposed to AdS/CFT (in which the CFT is said to exist at the boundary of AdS)? Constructing a logCFT correspondence in this scenario would be particularly interesting due to their non-unitary nature: the presence of hidden conformal symmetry has prompted many authors \cite{Castro:2010fd, Castro:2013kea,Haco:2018ske,Aggarwal:2019iay} to assume the validity of a Cardy formula and show that it reproduces the Bekenstein-Hawking entropy. However, Cardy's formula is not known to hold in non-unitary settings.
	
	This article is organized as follows. In Section \ref{sec:Hidden-symm-KG} we review how hidden conformal symmetry is found in the Klein-Gordon equation on a Kerr background.  In Section \ref{sec:standardform} we briefly present a standard form for the Klein-Gordon operator that will streamline our calculations. We note that this standard form holds for a probe field of general spin, in both four and five dimensions.  In Section \ref{sec:Hidden-symm-HD} we perform our analysis of hidden conformal symmetry in higher derivative dynamics. We calculate the monodromy data in Section \ref{sec:monodanal}, and move to holographic considerations in Section \ref{sec:HigherDHolography}. We discuss our findings and future work in Section \ref{sec:discussion}. In our appendix \ref{app:KGgoodform} we give more examples of the standard form Klein-Gordon operator presented in Section \ref{sec:standardform}.

	
	\section{Hidden conformal symmetry from the Klein-Gordon equation}
	\label{sec:Hidden-symm-KG}

	In this section we examine the hidden conformal symmetry of the Kerr black hole, as first discovered by~\cite{Castro:2010fd}. This will provide the framework for studying how hidden conformal symmetry is manifest in theories with higher derivative interactions in Section~\ref{sec:Hidden-symm-HD}. 
	While most of this section will be a review, some of the discussion we present is, we believe, new.
	
	The Kerr black hole is described by the following metric: 
	\begin{equation}\label{kerrMet}
		ds^2=\frac{\rho^2}{\Delta}dr^2-\frac{\Delta}{\rho^2}(dt^2-a~\text{sin}^2\theta d\phi)^2+\rho^2d\theta^2+\frac{\text{sin}^2\theta}{\rho^2}((r^2+a^2)d\phi-adt)^2,
	\end{equation}
	where we have defined 
	\begin{equation}
		\Delta=r^2+a^2-2 M r\,, \qquad \text{and}\qquad \rho^2=r^2+a^2\text{cos}^2\theta\,,
	\end{equation} and $a\equiv\frac{J}{M}$ is the spin of the black hole of mass $M$ and angular momentum $J$. This geometry possesses an outer horizon $r_+$ and inner horizon $r_-$, the locations of which are determined by the equation $\Delta=0$.
	In particular, we have
	\begin{equation}
		r_{\pm}= M\pm \sqrt{M^2-a^2}\,.
	\end{equation}
	There are two Killing vectors associated with (\ref{kerrMet}): $\partial_t$ and $\partial_{\phi}$. These generate explicit isometries of the metric (\ref{kerrMet}). 
	
	The additional hidden symmetry generators are not symmetries of the metric, but of the dynamics. Consider a massless scalar field $\Phi$ on the background (\ref{kerrMet}). Here we assume that we can treat the scalar $\Phi$ as a probe. 
	The Klein-Gordon equation of motion for this scalar, {\it i.e.}
	\begin{equation}
		{1\over \sqrt{g}}\partial_\mu(\sqrt{g} g^{\mu\nu}\partial_\nu\Phi)=0\,, \qquad \mu=0,\dots, 3\,,
	\end{equation}
	famously separates%
	\footnote{Indeed, the separability of this equation is actually a direct consequence of the existence of the hidden symmetry generators that we are about to build. For a review on this point see~\cite{Frolov:2006dqt, Frolov:2017kze}, and for a more recent discussion see~\cite{Keeler:2021tqy}.} 
	under the ansatz 
	\begin{equation}
		\Phi(t,r, \theta,\phi)=e^{i(m\phi-\omega t)}R(r)S(\theta)\,.
	\end{equation} 
	The radial equation is
	\begin{equation}\label{KleinGordon}
		\begin{split}
			&\left(\partial_r(\Delta\partial_r)+\frac{(\omega-\Omega_+m)^2}{4\kappa_+^2}\frac{(r_+-r_-)}{r-r_+}-\frac{(\omega-\Omega_-m)^2}{4\kappa_-^2}\frac{(r_+-r_-)}{r-r_-}\right.\\
			&\left.+(r^2+2M(r+2M))\omega^2\vphantom{\frac{1}{2}}\right)R(r)=KR(r),
		\end{split}
	\end{equation} 
	where $\Omega_{\pm}$ and $\kappa_{\pm}$ are the angular velocities and surface gravities of the inner and outer horizons, 
	\begin{equation}\label{omegas-kappas}
		\Omega_{\pm}=\frac{a}{2 M r_\pm} \,, \qquad \kappa_\pm=\frac{r_+-r_-}{4 M r_\pm}\,,
	\end{equation}
	and $K$ is a separation constant (which also encodes information on the spectrum of the spherical harmonic function $S(\theta)$). 
	
	\subsection{Near-region limit}
	\label{sec:near-region-lim}
	
	The authors of~\cite{Castro:2010fd, Haco:2018ske} have argued that a hidden conformal symmetry becomes manifest if we consider only soft hair modes. That is, they consider the following ``near-region'' limit of the equation \eqref{KleinGordon}: 
	\begin{equation}
		\omega \, M \ll 1\,,\qquad \text{and} \qquad \omega \, r \ll 1\,.
	\end{equation}
	As emphasized in \cite{Haco:2018ske}, this limit can be thought of as a near-horizon limit taken in phase space. The resulting equation is  
	\begin{equation}\label{KGlimit}
		\left(\partial_r(\Delta\partial_r)+\frac{(\omega-\Omega_+m)^2}{4\kappa_+^2}\frac{(r_+-r_-)}{r-r_+}-\frac{(\omega-\Omega_-m)^2}{4\kappa_-^2}\frac{(r_+-r_-)}{r-r_-}\right)R(r)=KR(r).
	\end{equation}
	
	The solutions of (\ref{KGlimit}) are hypergeometric functions. As pointed out by~\cite{Castro:2010fd}, the hypergeometric functions transform in representations of $\grSL(2,\R)$, which is the first hint of the existence of a hidden conformal symmetry. But what are the generators? To find them, we can take our inspiration from Kerr/CFT, and note that the Near-Horizon Extremal Kerr (NHEK) geometry is warped AdS$_3$, and for a particular choice of angle $\theta=\theta_0$ it is exactly the upper-half plane of AdS$_3$ (up to a conformal factor):
	\begin{equation}\label{AdSPoincare}
		ds^2=F(\theta_0)\left(\frac{dw^+dw^-+dy^2}{y^2}\right).
	\end{equation}
	Thus the existence of conformal symmetry of black horizons (hidden or otherwise) is tied to the existence of a copy of AdS$_3$ in the near-horizon limit (either in metric or in the Klein-Gordon equation).
	
	The isometry group of AdS$_3$ is $\grSL(2,\R)\times \grSL(2,\R)$, and we already know the Killing vectors for (\ref{AdSPoincare}). They are
	\begin{equation}\label{Hs}
		\begin{split}
			&H_1=i\partial_+, \qquad H_0=i\left(w^+\partial_++\frac{1}{2}y\partial_y\right),\qquad H_{-1}=i\left(w^{+2}\partial_++w^+y\partial_y-y^2\partial_-\right)\,,\\
			&\bar{H}_1=i\partial_-, \qquad \bar{H}_0=i\left(w^-\partial_-+\frac{1}{2}y\partial_y\right),\qquad \bar{H}_{-1}=i\left(w^{-2}\partial_-+w^-y\partial_y-y^2\partial_+\right).\\
		\end{split}
	\end{equation}
	These generators satisfy the conformal algebra
	\begin{equation}\label{commutationRels}
		\left[H_0,H_{\pm 1}\right]=\mp iH_{\pm}, \qquad \left[H_1,H_1\right]=-2iH_0\,,
	\end{equation}
	and have quadratic Casimir
	\begin{equation}\label{casimir}
		\begin{split}
			\mathcal{H}^2&=-H_0^2+\frac{1}{2}\left(H_1H_{-1}+H_{-1}H_1\right)\\
			&=\frac{1}{4}\left(y^2\partial_y^2-y\partial_y\right)+y^2\partial_+\partial_-.
		\end{split}
	\end{equation}
	
	Now the only thing that remains to identify the hidden symmetry generators of (\ref{KGlimit}) is to find a suitable coordinate transformation between Boyer-Lindquist coordinates $(t,r,\phi)$ and conformal coordinates $(w^{\pm},y)$. For Kerr black holes this turns out to be
	\begin{equation}\label{confcoor4d}
		\begin{split}
			w^+&=\left(\frac{r-r_+}{r-r_-}\right)^{1/2}e^{2\pi T_R\phi}\,,\\
			w^-&=\left(\frac{r-r_+}{r-r_-}\right)^{1/2}e^{2\pi T_L\phi-\frac{t}{2M}}\,,\\
			y&=\left(\frac{r_+-r_-}{r-r_-}\right)^{1/2}e^{\pi(T_L+T_R)\phi-\frac{t}{4M}}\,.
		\end{split}
	\end{equation}
	where
	\begin{equation}
		T_R=\frac{r_+-r_-}{4\pi a}, \qquad 	T_L=\frac{r_++r_-}{4\pi a}.
	\end{equation}
	Much has been written about the conformal coordinates (\ref{confcoor4d}) \cite{Castro:2010fd, Haco:2018ske, Aggarwal:2019iay, Perry:2020ndy}, and we will see more directly how to build them in Subsection \ref{subsec:Monodromy} regarding monodromy analysis\footnote{For instances of conformal coordinates in other contexts, see \cite{Maldacena:1998bw,Carlip:1995qv}.}. For now, we will just note that they are of the general form  
	\begin{equation}\label{confcoordgen}
		\begin{split}
			w^{+}&=f(r) e^{t_R},
			\\
			w^{-}&=f(r) e^{-t_L},
			\\
			y&=g(r)e^{(t_R-t_L)/2},
		\end{split}
	\end{equation}
	and we will return to $(t_L,t_R)$ later. It is important to note that plugging the coordinate transformation (\ref{confcoor4d}) into the Kerr metric (\ref{kerrMet}) does not reproduce the Poincar\'e patch metric (\ref{AdSPoincare}) exactly. Rather, near the bifurcation surface $w^{\pm}=0$, the Kerr metric becomes
	\begin{equation}\label{metexpand}
		ds^2=\frac{4\rho^2_+}{y^2}dw^+dw^-+\frac{16J^2\sin^2\theta_0}{y^2\rho^2_+}dy^2+\rho_+^2d\theta^2+\dots\,,
	\end{equation}
	where 
	\begin{equation}
		\rho_+^2=r_+^2+a^2\cos^2\theta\,,
	\end{equation}
	and the terms in the ellipsis ``$\dots$'' in equation \eqref{metexpand} are at least linear order in the coordinates $w^{\pm}$. The existence of these higher order terms underscores the fact that the hidden conformal symmetry generators (\ref{Hs}) are not isometries of the Kerr metric \eqref{kerrMet}. 
	
	For clarity, expressions of the generators (\ref{Hs}) in Boyer-Lindquist coordinates are~\cite{Castro:2010fd}
	\begin{equation}\label{BLgens}
		\begin{split}
			H_1&=ie^{-2\pi T_R\phi}\left(\Delta^{1/2}\partial_r+\frac{1}{2\pi T_R}\frac{r-M}{\Delta^{1/2}}\partial_{\phi}+\frac{2T_L}{T_R}\frac{Mr-a^2}{\Delta^{1/2}}\partial_t\right),\\
			H_0&=\frac{i}{2\pi T_R}\partial_{\phi}+2iM\frac{T_L}{T_R}\partial_t,\\
			H_{-1}&=ie^{2\pi T_R\phi}\left(-\Delta^{1/2}\partial_r+\frac{1}{2\pi T_R}\frac{r-M}{\Delta^{1/2}}\partial_{\phi}+\frac{2T_L}{T_R}\frac{Mr-a^2}{\Delta^{1/2}}\partial_t\right),\\
		\end{split}
	\end{equation} 
	with similar expressions for the $\bar{H}$ sector. In these coordinates, the quadratic Casimir (\ref{casimir}) becomes exactly the near-region radial Klein-Gordon operator in (\ref{KGlimit}), so that
	\begin{equation}\label{HPhiKPhi}
		\mathcal{H}^2\Phi=K\Phi.
	\end{equation}
	
	\subsection{Monodromy Method}
	\label{subsec:Monodromy}
	
	We will now see that it is possible to find the generators (\ref{BLgens}) without explicitly taking a ``near-region'' limit, as in~\cite{Castro:2010fd}. This subsection will largely follow~\cite{Aggarwal:2019iay}, with some new discussion. 
	
	Let's again consider the Klein-Gordon equation (\ref{KleinGordon}). This differential equation has two regular singular points at the horizons $r_{\pm}$, and one irregular singular point at infinity.%
	\footnote{Our discussion will only require the regular singular points. For treatment of the irregular singular point see~\cite{Castro:2013kea, Castro:2013lba}.} 
	Each singular point causes a branch cut, and we are interested in studying the radial solutions $R(r)$ of (\ref{KleinGordon}) (now promoted to complex-valued functions) when we go around each of the regular singular points. In general, the solutions $R(r)$ will develop a monodromy around these singular points. To study this, we posit that $R(r)$ has a series solution of the form 
	\begin{equation}\label{seriessol}
		R(r)=(r-r_i)^{\beta}\sum_{n=0}^{\infty}q_n(r-r_i)^n.
	\end{equation}
	
	Our immediate objective is to determine the monodromy parameter $\beta\equiv i\alpha$ using the Frobenius method. We will go through this in more detail for our more complicated higher derivative case in Section \ref{sec:Hidden-symm-HD}. For the case at hand, we just state the answer and refer the reader to~\cite{Castro:2013kea} for details. The monodromy parameters around the inner and outer horizons are
	\begin{equation}\label{alphas}
		\alpha_{\pm}=\frac{\omega-\Omega_{\pm}m}{2\kappa_{\pm}},
	\end{equation}	
	where $\Omega_\pm$ and $\kappa_\pm$ where defined in \eqref{omegas-kappas}. 
	
	Next, a crucial step in obtaining the generators (\ref{BLgens}) is to implement a change of basis
	\begin{equation}\label{expbasis}
		e^{i(m\phi-\omega t)}=e^{-i(\omega_Lt_L+\omega_Rt_R)}.
	\end{equation}	
	The choice employed by~\cite{Aggarwal:2019iay} is
	\begin{equation}\label{omegabasis}
		\omega_L=\alpha_+-\alpha_-, \qquad \omega_R=\alpha_++\alpha_-.
	\end{equation}
	The particular change of basis (\ref{omegabasis}) is not well-motivated in the literature, and is often taken as a purely mathematical step to match the results of~\cite{Castro:2010fd}. To determine whether this is the appropriate basis choice to use in our more complicated higher derivative analysis in Section \ref{sec:Hidden-symm-HD}, a deeper physical understanding of this choice is needed, which we outline here\footnote{Key elements of this discussion were also presented in \cite{Castro:2013kea}.}. 
	
	Let's consider how the radial solutions $R(r)$ change as we go around the singular point $r_+$. Near $r=r_+$, our radial solutions are of the form 
	\begin{equation}
		R(r)=(r-r_+)^{\pm i\alpha_+}\left(1+\mathcal O(r-r_+)\right)\,.
	\end{equation}
	When we go around the singular point, $r-r_+\rightarrow e^{2\pi i}(r-r_+)$, we see that 
	\begin{equation}
		R(r)\rightarrow R(r)e^{\mp 2\pi\alpha_+}\,. 
	\end{equation}
	As explained in \cite{Castro:2013kea}, if we go around the singular point $r_+$ twice, i.e. $r-r_+\rightarrow e^{4\pi i}(r-r_+)$, we expect the wave equation $\Phi=e^{-i\omega t+im\phi}R(r)$  to be invariant. 
	The radial piece of the outgoing solution $R(r)=(r-r_+)^{i\alpha_+}$ picks up the factor $e^{-4\pi\alpha_+}$. Plugging in the value for $\alpha_+$ in (\ref{alphas}), we find that $t$ and $\phi$ must transform in such a way to cancel this factor:
	\begin{equation}\label{transform}
		(t,\phi)\sim(t,\phi)+\frac{2\pi i}{\kappa_+}(1,\Omega_+).
	\end{equation}
	
	We will now see that the basis choice (\ref{omegabasis}) arises from determining the appropriate conjugate variables that will lead to the more natural thermal and angular transformation properties. For example, around $r=r_+$ we would like to replace (\ref{transform}) with
	\begin{equation}\label{betterTransf}
		(X,Y)\sim(X,Y)+2\pi i(1,1).
	\end{equation}
	The authors of \cite{Castro:2013kea} presciently rename $(X, Y)$ as $(t_L, t_R)$. 
	Similar arguments for the singular point $r_-$ lead us to the transformation properties
	\begin{equation}\label{betterTransf-minus}
		(X,Y)\sim(X,Y)+2\pi i(-1,1).
	\end{equation}
For the wavefunction $\Phi=e^{-i\omega_Lt_L-i\omega_R t_R}R(r)$, we find by using (\ref{betterTransf}) and \eqref{betterTransf-minus}, 
	the functions $(t_L, t_R)$ that accomplish these identifications are 
	\begin{equation}\label{tpm}
		t_R=2\pi T_R\phi\,, \qquad t_L=\frac{1}{2M}t-2\pi T_L\phi\,.
	\end{equation}	
	
	With $(t_L, t_R)$ in (\ref{tpm}), we can now immediately reproduce the zero-mode generators of~\cite{Castro:2010fd}. They are
	\begin{equation}
		H_0=\frac{i}{2\pi T_R}\partial_{\phi}+2iM\frac{T_L}{T_R}\partial_t=i\partial_{t_R}, \qquad \bar{H}_0=-2iM\partial_t=-i\partial_{t_L}.
	\end{equation}
	Notice that in this discussion we avoided making the seemingly arbitrary basis choice (\ref{omegabasis}). Instead, we see that $(\omega_L,\omega_R)$ are fixed by (\ref{tpm}) and (\ref{expbasis}). 
	
	It might appear that the monodromy analysis only determines the zero-mode generators $(H_0,\bar{H}_0)$ and not $(H_{\pm 1},\bar{H}_{\pm 1})$, but we can actually go further. From equation \eqref{confcoordgen}, we see that $(t_L, t_R)$ also fix the conformal coordinates (up to a radial factors, which can be recovered from the radial behavior of the hypergeometric solutions of \eqref{KGlimit} or from the Klein-Gordon operator itself, which we argue in Section \ref{sec:Hidden-symm-HD} and Appendix \ref{app:KGgoodform}). Once we have the proper conformal coordinates, all of the $H$s are determined by \eqref{Hs}.

	\section{A standard form for the Klein-Gordon operator}\label{sec:standardform}
	Before we move on to our higher derivative model, it is useful for us to express the Klein-Gordon operator in a standard form. In addition to streamlining our analysis, this form highlights interesting physical structure of the operator, that persists in both higher dimensional and higher spin settings.
	
	We begin by writing the Klein-Gordon operator on the Kerr background in the following way:
	\begin{equation}
		\begin{split}
			\nabla^{\mu}\nabla_{\mu}&=\frac{1}{\rho^2}\left[\partial_r(\Delta\partial_r)-\frac{(r_+-r_-)}{(r-r_+)}\left(\frac{\partial_t+\Omega_+\partial_{\phi}}{2\kappa_+}\right)^2+\frac{(r_+-r_-)}{(r-r_-)}\left(\frac{\partial_t+\Omega_-\partial_{\phi}}{2\kappa_-}\right)^2\right.\\
			&\left.+\frac{1}{\sin^2\theta}\partial_{\phi}^2-(a^2\cos^2\theta+4M^2)\partial_t^2-(r^2+2Mr)\partial_t^2+\frac{1}{\sin^2\theta}\partial_{\theta}\left(\sin^2\theta\partial_{\theta}\right)\right].
		\end{split}
	\end{equation}
	This form has several useful features for our analysis. First, if we posit the standard solution $\Phi=e^{i(m\phi-\omega t)}R(r)S(\theta)$, we see that the term
	\begin{equation}\label{CarterOp}
		\left(\frac{1}{\sin^2\theta}\partial_{\phi}^2-(a^2\cos^2\theta+4M^2)\partial_t^2\right)\Phi=C\Phi
	\end{equation}
	produces a version of Carter's constant $C$ \cite{carter1968global}. This means that, if we consider a constant $\theta$ slice $\theta=\theta_0$, the only dependence on our choice of slice will be in the prefactor $\rho^{-2}$ (recall that $\rho^2=r^2+a^2\cos^2\theta$). The presence of this factor means that our higher order equation of motion $\left(	\nabla^{\mu}\nabla_{\mu}\right)^n\Phi=0$ appears not to be separable. Nevertheless, we will see in Section \ref{sec:Hidden-symm-HD} that at leading order near $r=r_\pm$, all dependence on $\rho^2$ (and thus $\theta_0$) will drop out. Thus we consider a constant $\theta$ slice $\theta=\theta_0$, allowing us to study the ``radial'' operator 
	\begin{equation}\label{radialKG}
		\begin{split}
			\nabla^{\mu}\nabla_{\mu}&=\frac{1}{\rho_0^2}\left[\partial_r(\Delta\partial_r)-\frac{(r_+-r_-)}{(r-r_+)}\left(\frac{\partial_t+\Omega_+\partial_{\phi}}{2\kappa_+}\right)^2+\frac{(r_+-r_-)}{(r-r_-)}\left(\frac{\partial_t+\Omega_-\partial_{\phi}}{2\kappa_-}\right)^2\right.\\
			&\left.C_{t\phi}-(r^2+2Mr)\partial_t^2\vphantom{\frac{1}{1}}\right],
		\end{split}
	\end{equation}
	where $\rho_0^2=r^2+a^2\cos^2\theta_0$ and we have called the operator in \eqref{CarterOp} $C_{t\phi}$ for convenience.

	The form of the operator \eqref{radialKG} has a further use. The terms
	\begin{equation}
		\frac{\partial_t+\Omega_\pm\partial_{\phi}}{2\kappa_\pm}
	\end{equation} give the monodromy parameters \eqref{alphas} introduced in Section \ref{subsec:Monodromy}, that is,
	\begin{equation}
		\left(\frac{\partial_t+\Omega_\pm\partial_{\phi}}{2\kappa_\pm}\right)\Phi=-i\alpha_{\pm}\Phi\,.
	\end{equation}
	In addition, the Killing vector fields
	\begin{equation}
		\xi_\pm=\kappa_{\pm}\left(\partial_t+\Omega_\pm\partial_{\phi}\right) 
	\end{equation}
	are exactly those that vanish on the inner and outer horizons $r_{\pm}$. As such, they are the same vector fields that appear in Wald's formulation \cite{Wald:1993nt} of black hole entropy as the integrated Noether charge associated with the Killing vectors vanishing on the horizons. This point was discussed in \cite{Castro:2013kea}. 
	
	Further, the radial factors $\left(\frac{r_+-r_-}{r-r_\pm}\right)$ can be directly related to the conformal coordinates as defined in \eqref{confcoordgen}, and we can finally write the Klein-Gordon operator \eqref{radialKG} acting on $\Phi$ as
	\begin{equation}\label{radialKG-Phi}
		\nabla^{\mu}\nabla_{\mu}\Phi=\frac{\Phi}{\rho^2R(r)}	\left[\partial_r(\Delta\partial_r)+\alpha_+^2\frac{g^2(r)}{f^2(r)}-\alpha_-^2g^2(r)+(r^2+2Mr)\omega^2+C_{t\phi}\right]R(r).
	\end{equation} 
	This form of the Klein-Gordon operator holds for higher spin fields \textit{and} in higher dimensions, even though the forms of $\alpha_{\pm}$ and $f(r)$ and $g(r)$ change. This is discussed in Appendix \ref{app:KGgoodform}. The only terms that do change for higher spin/higher dimension are the non-singular terms 
	\begin{equation}
		(r^2+2Mr)\omega^2+C_{t\phi},
	\end{equation}
	which are precisely those that are dropped in the near-region limit of Section \ref{sec:near-region-lim}.
	
	\section{Hidden conformal symmetry in higher derivative dynamics}
	\label{sec:Hidden-symm-HD}
	
	To study how hidden conformal symmetry is manifest in a theory with higher derivative dynamics, we consider the following action for a massless scalar field on a Kerr black hole background
	\begin{equation}\label{action}
		S=-\int d^{4}x\sqrt{g}\left(\frac{1}{2}\Phi\left(\nabla^{\mu}\nabla_{\mu}\right)^n\Phi\right),
	\end{equation}
	where $n$ is an integer. The equation of motion resulting from this action is
	\begin{equation}\label{genreq}
		\left(	\nabla^{\mu}\nabla_{\mu}\right)^n\Phi=0,
	\end{equation}
	where $n=1$ is the Klein-Gordon equation. Our motivation for choosing this action is twofold. First, the equations of motion are simple enough as to provide a straightforward extension to previous results with $n=1$ obtained by \cite{Castro:2010fd} and \cite{Aggarwal:2019iay}. Though simple, we will see that \eqref{genreq} already provides interesting complications that give insight into whether the choice of dynamics affects hidden conformal symmetry. Our second motivation is that the action \eqref{action} is of physical interest, since known examples of holographic duals to logarithmic conformal field theories contain higher derivative equations of motion \cite{Hogervorst:2016itc}. Thus, \eqref{action} provides us with the opportunity to both study the effect of changing the dynamics on hidden conformal symmetry, while also potentially diagnosing a new instance of a logCFT correspondence. 
	
	The differential equation \eqref{genreq} can be reformulated in two ways, one of which will be of particular use to us. The aim of both approaches is to reduce the system to a series of second order equations. For example, as was discussed in \cite{Bergshoeff:2012sc}, the equation of motion \eqref{genreq} can be broken up into coupled second order equations by introducing $n-1$ auxiliary scalar fields
	\begin{equation}\label{auxfields}
		\begin{split}
			&\nabla_\mu\nabla^\mu\Phi_1=0,\\
			&\nabla_\mu\nabla^\mu\Phi_i=\Phi_{i-1}, \qquad \text{for}~ i=2, ...,n.\\
		\end{split}
	\end{equation}
	The related and more useful alternative to this approach is to repackage the auxiliary scalar fields as higher spin objects. That is, the problem \eqref{genreq} can be expressed as 
	\begin{equation}\label{HigherSpinEq}
		\begin{split}
			&	\nabla^{\mu}\nabla_{\mu}\Phi=0,\\
			&	\nabla^{\mu}\nabla_{\mu}\Phi_{\mu_1\mu_2}=0, \\
			&	\nabla^{\mu}\nabla_{\mu}\Phi_{\mu_1\mu_2\mu_3\mu_4}=0, \\
			&\vdots\\
			&
			\nabla^{\mu}\nabla_{\mu}\Phi_{\mu_1...\mu_{2n-2}}=0.
		\end{split}
	\end{equation}
	In the above expressions we have defined defined the higher spin fields as
	\begin{equation}\label{HigherSpinDefs}
		\begin{split}
			&\Phi_{\mu_1\mu_2}\equiv\nabla_{\mu_1}\nabla_{\mu_2}\Phi,\\ &\Phi_{\mu_1\mu_2\mu_3\mu_4}\equiv\nabla_{\mu_1}\nabla_{\mu_2}\nabla_{\mu_3}\nabla_{\mu_4}\Phi, \\
			& \vdots\\
			&\Phi_{\mu_1\mu_2...\mu_{2n-2}}\equiv\nabla_{\mu_1}\nabla_{\mu_2}...\nabla_{\mu_{2n-2}}\Phi.\\
		\end{split}
	\end{equation}
	We will return to the significance of this reformulation later in Section  \ref{sec:modelinads}. 
	
	We begin this section with a monodromy analysis of higher derivative dynamics on a Kerr background in Section \ref{sec:monodanal}. As mentioned before, one initial motivation for studying higher derivative dynamics is to potentially identify a new instance of a logCFT correspondence through hidden conformal symmetry, in the spirit of the Kerr/CFT correspondence away from extremality. We study this question in Section \ref{sec:modelinads}, which we begin with a short review of how the higher derivative model \eqref{action} is used in AdS/logCFT, followed by a study of whether this is a viable model with which to build a Kerr/logCFT correspondence.


	\subsection{Monodromy analysis}\label{sec:monodanal}
	In this section we analyze hidden conformal symmetry in our higher derivative theories in the spirit of Section \ref{subsec:Monodromy}.  We begin our analysis with $n=2$ in the equation of motion \eqref{genreq}. We will then treat $n=3$, and construct a clear pattern for the monodromy parameters $\alpha_{\pm}$ for general $n$.
	
	We immediately encounter the would-be issue that for $n>1$ the equation of motion \eqref{genreq} appears not to be separable. Perhaps intriguingly, this turns out not to matter at leading order near $r=r_\pm$. That is, in what follows, we take a constant slice $\theta=\theta_0$, and our results for $\alpha_\pm$ do not depend on the choice of $\theta_0$.

	\subsubsection{Case $n=2$}
	Since we are free to take a constant $\theta$ slice, we can focus on the behavior of a radial  differential equation near its singular points. In particular, we can write the radial equation in standard form near a singular point $r=r_i$:
	\begin{equation}\label{fourthdiffeq}
		(r-r_i)^4R^{(4)}+D(r)(r-r_i)^3R^{(3)}+C(r)(r-r_i)^2R''+B(r)(r-r_i)R'+A(r)R=0.
	\end{equation}
	The Frobenius method instructs us to look for series solutions of the form
	\begin{equation}\label{series}
		\begin{split}
			R(r)&=(r-r_i)^{\beta}\sum_{k=0}^{\infty}q_k(r-r_i)^k, 
		\end{split}
	\end{equation}
	and coefficient functions expanded as
	\begin{equation}\label{coefseries}
		\begin{split}
			D(r)&=\sum_{k=0}^{\infty}d_k(r-r_i)^k,\qquad C(r)=\sum_{k=0}^{\infty}c_k(r-r_i)^k,\\ B(r)&=\sum_{k=0}^{\infty}b_k(r-r_i)^k, \qquad A(r)=\sum_{k=0}^{\infty}a_k(r-r_i)^k.
		\end{split}
	\end{equation}
	In order for (\ref{fourthdiffeq}) to be satisfied, the coefficient of each power of $r-r_i$ must equal zero. In particular, the coefficient of the $(r-r_i)^{\beta}$ term gives us the fourth order indicial equation
	\begin{equation}
		\beta(\beta-1)(\beta-2)(\beta-3)+\beta(\beta-1)(\beta-2)d_0+\beta(\beta-1)c_0+\beta b_0+a_0=0.
	\end{equation}
	
	Without loss of generality, we first study the analytic structure around $r=r_+$. The zeroth order coefficients of our series expansions (\ref{series}) are
	\begin{equation}
		\begin{split}
			d_0=4,\qquad c_0=2(1+\alpha^2_+),\qquad b_0=0, \qquad a_0=\alpha^2_+(1+\alpha^2_+),
		\end{split}
	\end{equation}
	where $\alpha_+$ denotes the monodromy parameter for a scalar field on a Kerr blackground near the outer horizon, as defined in \eqref{alphas}. Plugging these values back into our indicial equation, we get
	\begin{equation}
		(\beta^2+\alpha^2_+)((\beta-1)^2+\alpha^2_+)=0.
	\end{equation}
	Thus, we obtain
	\begin{equation}\label{betanequals2}
		\beta=\left\{\pm i\alpha_+,~1\pm i\alpha_+\right\},
	\end{equation}
	where $\alpha_+$ is given by the expression \eqref{alphas}, 
	or, since $\beta\equiv i\alpha$,
	\begin{equation}\label{alphapn2}
		\alpha^{n=2}_+=\left\{\pm\alpha_+,~-i\pm\alpha_+\right\}.
	\end{equation}
	There is a similar result for $\alpha^{n=2}_-$:
	\begin{equation}\label{alphaminn2}
		\alpha^{n=2}_-=\left\{\pm\alpha_-,~-i\pm\alpha_-\right\}\,,
	\end{equation}
	where again $\alpha_-$ can be read in equation \eqref{alphas}. 
	
	At this point there are several things to point out regarding \eqref{betanequals2}-\eqref{alphaminn2}. First we can see why it is useful to reformulate our equation of motion \eqref{genreq} for $n=2$ as two coupled equations with a higher spin field, as in \eqref{HigherSpinEq}:
	\begin{equation}\label{HigherSpinn2}
		\begin{split}
			&	\nabla^{\mu}\nabla_{\mu}\Phi=0,\\
			&	\nabla^{\mu}\nabla_{\mu}\Phi_{\mu_1\mu_2}=0. \\
		\end{split}
	\end{equation}
	These equations are coupled in the sense that $\Phi_{\mu_1\mu_2}$ is built from $\Phi$ as in \eqref{HigherSpinDefs}. The four monodromy parameters $\alpha^{n=2}_+$ associated with our fourth-order equation near the outer horizon $r_+$ are exactly those that were found when analyzing the second-order Klein-Gordon equation for a scalar field ($\alpha_\pm$, see equation \eqref{alphas}) and for a spin-2 field ($-i\pm\alpha_\pm$, see equation \eqref{alphaspins}). 
	
	Second, from \eqref{betanequals2} we see that two of the exponents $\beta$ differ from two others by a positive integer. The Frobenius method tells us that of the four linearly independent solutions around each singular point, two of them \textit{might be} log solutions. For example, near $r=r_+$ we have
	\begin{equation}
		\begin{split}
			R(r)&=(r-r_+)^{1\pm i\alpha_+}\Phi_1^{\pm}(r), \\
			R(r)&=a_{\pm}(r-r_+)^{1\pm i\alpha_+}\Phi_1^{\pm}(r)\log (r-r_+)+(r-r_+)^{\pm i\alpha_+}\Phi_2^{\pm}(r),
		\end{split}
	\end{equation}
	where $a_{\pm}$ are constants which can be zero or not. This could signal that, if there is indeed a CFT description of this system, it could be a logCFT. However, it is important to note that there is a subtle difference between the log terms that appear here for Kerr and those which appear in the context of logCFTs dual to an AdS background, as discussed in Section \ref{sec:modelinads}. In the Frobenius method, when two roots are repeated (as in the AdS analysis, see equation \eqref{multiplicity}), a logarithmic part of the solution is guaranteed. In contrast, it is a theorem that when two roots differ by a positive integer (as in our case) the coefficients $a_{\pm}$ could be zero, see {\it e.g.}~\cite{Coddington1955TheoryOO}. This depends on the specific and intricate nature of the given differential equation.

	Our principal goal is now to study if and how hidden conformal symmetry is manifest in our higher derivative dynamics $\left(\nabla_\mu\nabla^\mu\right)^2\Phi=0$. There are several ways to approach this problem. In the usual scenario, that is $\nabla_\mu\nabla^\mu\Phi=0$, we try to find $\grSL(2,\mathbb R)$ generators $(H_0,H_{\pm 1})$ as in~\eqref{BLgens} that 1) satisfy the commutation relations \eqref{commutationRels} and 2) form a Casimir that reproduces the near-region Laplacian $\mathcal{H}^2\Phi=K\Phi$, as in \eqref{HPhiKPhi}. 
	Trying to find equivalent structure in the equation $\left(\nabla_\mu\nabla^\mu\right)^2\Phi=0$ directly suffers from conceptual issues, as it appears the role of a quadratic Casimir $\mathcal{H}^2$ would perhaps have to be replaced by a quartic Casimir $\mathcal{H}^4$.
	However, if we take the equivalent description of our system \eqref{HigherSpinn2}, we will see that hidden conformal symmetry is still visible, and presents itself in a natural way.

	Since \eqref{HigherSpinn2} is an equivalent description of our fourth order equation, we can analyze each equation in \eqref{HigherSpinn2} separately. The first equation, $\nabla^{\mu}\nabla_{\mu}\Phi=0$, is of course just the standard case that was already treated in \cite{Aggarwal:2019iay}. Now we turn to the spin-2 equation. For the reader's convenience, we reproduce this equation here from our Appendix \ref{app:KGgoodform}:
	\begin{equation}
		\left(\partial_r\Delta\partial_r+\alpha^2_{+, s=2}\frac{g^2(r)}{f^2(r)}-\alpha^2_{-, s=2}g^2(r)+\omega^2r^2+2(M\omega+2i)\omega r+C_{t,\phi}\right)R(r)=0,
	\end{equation}
	where $s$ is the spin of the auxiliary field. Again, the constant $C_{t,\phi}$ we can think of as being absorbed in a separation constant, and the terms $\omega^2r^2+2(M\omega+2i)\omega r$ can be dropped in the near-region limit. Thus the solutions to 
	\begin{equation}
		\left(\partial_r\Delta\partial_r+\alpha^2_{+, s=2}\frac{g^2(r)}{f^2(r)}-\alpha^2_{-, s=2}g^2(r)\right)R(r)=0
	\end{equation}
	are also hypergeometric functions, hinting at hidden conformal symmetry. 
	
	In our review Section \ref{sec:Hidden-symm-KG} we introduced several important and interrelated quantities: monodromy exponents $\beta\equiv i\alpha$~\eqref{alphas}, the change of basis modes $(\omega_L, \omega_R)$~\eqref{omegabasis}, their conjugate variables $(t_L, t_R)$~\eqref{tpm}, the conformal coordinates $(w^\pm,y)$~\eqref{confcoor4d}, and the $\grSL(2,\mathbb R)$ generators $(H_0,H_{\pm 1})$~\eqref{BLgens}. We now ask the question: which of these quantities, if any, need to be modified from their $n=1$ values so that we can still obtain the conditions for diagnosing hidden conformal symmetry? Notice that this is equivalent to finding generators that satisfy the commutation relations \eqref{commutationRels} whose Casimir reproduces the near-region Klein-Gordon operator $\mathcal{H}^2_{n=2}\Phi_{\mu_1\mu_2}=K\Phi_{\mu_1\mu_2}$, as in \eqref{HPhiKPhi}. We claim that this is accomplished by modifying only one thing: the change of basis choice $(\omega_L, \omega_R)$. Let's see how this works by discussing each of the above quantities in turn. 
	
	First, we claim that $(t_L, t_R)$ do not change from their $n=1$ values given in~\eqref{tpm}, since these were obtained by thermal considerations in our review Section \ref{sec:Hidden-symm-KG}. The conformal coordinates $(w^\pm,y)$ should also not change from \eqref{confcoor4d} and \eqref{confcoordgen}, since these are just purely geometric relations taking the Kerr background to the upper-half plane (to leading order near the bifurcation surface). Finally, the generators are built directly from the the conformal coordinates via \eqref{Hs}, so these should also remain unchanged from their $n=1$ values. This only leaves two quantities: the monodromy parameters $\alpha$, which certainly \textit{do} change, and the basis choice $(\omega_L, \omega_R)$, which must change also to account for the change in the $\alpha$s. As mentioned above and in Appendix \ref{app:KGgoodform}, the new $\alpha$s for the spin-2 equation are 
	\begin{equation}
		\alpha^{s=2}_\pm=i\pm\alpha^{s=0}_\pm,
	\end{equation}
	and the basis choice to accommodate this is modified from \eqref{omegabasis} to
	\begin{equation}\label{newbasischoice}
		\omega_L=\alpha_+-\alpha_--2i, \qquad \omega_R=\alpha_++\alpha_-.
	\end{equation}
	This means that we consider the frequencies $\omega\in\mathbb{C}$. 
	
 It is not illuminating to write the full equation $\left(\nabla_\mu\nabla^\mu\right)^2\Phi=0$ in a standard form (as in Section \ref{sec:standardform}) except to point out one thing. Taking a constant $\theta$ slice, the fourth order equation of motion is of the form:
	\begin{equation}\label{standardhighern}
		\mathcal{D}\left[R(r)\right]+\left[\alpha_+^2\left(1+\alpha_+^2\right)\frac{g^4(r)}{f^4(r)}+\alpha_-^2\left(1+\alpha_-^2\right)g^4(r)+n.s.\right]R(r)=0,
	\end{equation}
	where $\mathcal{D}\left[R(r)\right]$ stands for all terms involving a derivative of $R(r)$, $n.s.$ represents nonsingular terms, and the radial functions $f$ and $g$ are defined as in \eqref{confcoor4d} and \eqref{confcoordgen}. From equation \eqref{standardhighern} we learn more about the standard form discussed in Section \ref{sec:standardform}: the coefficients of the radial functions are just the Frobenius exponents $\beta$ (see equation \eqref{betanequals2}).

	\subsubsection{Case $n=3$ and higher $n$}
	We now sketch the monodromy calculation for the equation of motion $\left(\nabla^\mu\nabla_\mu\right)^n\Phi=0$ with $n=3$, and establish a pattern for general $n$. 
	
	Just as in the $n=2$ case, our analysis does not depend upon our choice of constant $\theta$ slice. We are thus free to consider the sixth order radial equation in standard form
	\begin{equation}
		\begin{split}
			&(r-r_i)^6R^{(6)}+F(r)(r-r_i)^5R^{(5)}+E(r)(r-r_i)^4R^{(4)}+\\&D(r)(r-r_i)^3R^{(3)}
			+C(r)(r-r_i)^2R''+B(r)(r-r_i)R'+A(r)R=0.
		\end{split}
	\end{equation}
	For concreteness, we again choose to study $r_i=r_+$. Upon positing a series solution for $R(r)$ and the coefficient functions (as in \eqref{series} and \eqref{coefseries}), we find  the indicial equation
	\begin{equation}
		(\beta^2+\alpha_+^2)((\beta-1)^2+\alpha_+^2)((\beta-2)^2+\alpha_+^2)=0.
	\end{equation}
	So all together our indicial equations close to $r=r_+$ for $\left(\nabla_\mu\nabla^\mu\right)\Phi=0$, $\left(\nabla_\mu\nabla^\mu\right)^2\Phi=0$ and $\left(\nabla_\mu\nabla^\mu\right)^3\Phi=0$ are
	\begin{equation}\label{allindicials}
		\begin{split}
			&(\beta^2+\alpha_+^2)=0,\\
			&(\beta^2+\alpha_+^2)((\beta-1)^2+\alpha_+^2)=0,\\
			&(\beta^2+\alpha_+^2)((\beta-1)^2+\alpha_+^2)((\beta-2)^2+\alpha_+^2)=0,
		\end{split}
	\end{equation}
	respectively. 
	Equation \eqref{allindicials} suggests the monodromy structure of $\left(\nabla^\mu\nabla_\mu\right)^n\Phi=0$ on the Kerr background for general $n$:
	\begin{equation}
	\prod_{j=1}^{n}((\beta-j+1)^2+\alpha_+^2)=0.
	\end{equation}
	
	\subsection{Holographic correspondences with higher derivative dynamics}\label{sec:HigherDHolography}
	
		\subsubsection{AdS/logCFT}\label{sec:modelinads}
	
	Here we begin with a brief review of how higher derivative dynamics in AdS are dual to logCFTs, in order to contrast what happens when we consider these dynamics on a Kerr background in the next subsection. Holographic logCFTs have been discussed since the early days of AdS/CFT~\cite{Ghezelbash:1998rj,Kogan:1999fo}. There has been much progress in this direction, see {\it e.g.} \cite{Hogervorst:2016itc}, and here we only report the main lesson from it: logCFTs are holographically realised as higher derivative theories in AdS$_{d+1}$ spacetimes. In particular, for scalar fields with mass $\mu$ the action is
	\begin{equation}\label{action-laplacian-n}
		S=-\frac 12 \int \dif^{\, d+1}x \sqrt g\, \phi\left(\nabla_\mu\nabla^\mu-\mu^2\right)^n\phi, 
	\end{equation}
	where $n\ge 2$ corresponds to the rank $n$ of the dual logCFT. 
	The equation of motion is then a $2n$-th order differential equation, {\it i.e.} 
	\begin{equation}\label{eom-laplacian-n}
		\left(\nabla_\mu\nabla^\mu-\mu^2\right)^n\phi=0\,.
	\end{equation}
	The above action can also be formulated in terms of auxiliary fields, for example in the case $n=2$ we have
	\begin{equation}\label{action-n2-aux}
		S=-\frac 12 \int \dif^{\,d+1}x \sqrt g \left( g^{\mu\nu} \partial_\mu \phi_1 \partial_\nu \phi_2+ \mu^2\phi_1\phi_2+\frac 12 \phi_1^2\right)\,.
	\end{equation}
	The equation of motion for $\phi_1 $ and $\phi_2$ are respectively
	\begin{equation}
		\left(\nabla_\mu\nabla^\mu-\mu^2\right)\phi_2=\phi_1\,, \qquad
		\left(\nabla_\mu\nabla^\mu-\mu^2\right)\phi_1=0\,. \\
	\end{equation}
	It is then clear that the $\phi_2$ has to satisfy a ``squared'' equation, that is $\left(\nabla_\mu\nabla^\mu-\mu^2\right)^2\phi_2=0$. The action \eqref{action-n2-aux} can be generalised to arbitrary rank $n$~\cite{Bergshoeff:2012bi}, and the corresponding equations of motion are given by
	\begin{align}
		& \left(\nabla_\mu\nabla^\mu-\mu^2\right)\phi_1=0\,,
		\\ \nonumber
		&  \left(\nabla_\mu\nabla^\mu-\mu^2\right)\phi_{i}=\phi_{i-1}\,,\qquad i=2, \dots, n\,,
	\end{align}
	from which follows that the equation of motion for the $n$-th field is indeed \eqref{eom-laplacian-n}.
	In terms of the auxiliary fields it is manifest a shift symmetry of the equations of motion and the on-shell action
	\begin{equation}
		\phi_i \to \phi_i +\sum_{p=1}^{i-1}\lambda_{p} \phi_p\,,
	\end{equation}
	for arbitrary constant $\lambda_{p}$. 



	Here we briefly illustrate the example of a higher rank wave equation for a scalar field in (Euclidean) AdS$_{d+1}$; see {\it e.g.} \cite{Hogervorst:2016itc} for a recent review. We assume the scalar field to be massless, and we use a Poincar\'e patch, that is 
	\begin{equation}
		\label{ads-metric}
		ds^2_{AdS}=\frac{\dif \zeta^2+\dif x_i \dif x^i}{\zeta^2}\,,
	\end{equation}
	where $i=1, \dots, d$. 
	
	We are looking at the equation \eqref{genreq} for the case $n=2$ and in the background \eqref{ads-metric}. Given the high degree of symmetry of the metric \eqref{ads-metric} the differential equation is particularly simple. We can take $d=2$ to make a direct comparison with the Kerr black hole metric \eqref{kerrMet} discussed in this work. 
	Then, the radial differential equation for the radial component $\psi(\zeta)$ of the scalar field  is given by 
	\begin{equation}\label{radialODE-ads4}
		\psi^{(4)}(\zeta)
		+{2\over \zeta} \psi^{(3)}(\zeta)
		-2\left( \mathcal{K}^2+\frac{1}{\zeta^2}\right) \psi^{(2)}(\zeta)
		+\left(\frac{1}{\zeta^3}-{2 \mathcal{K}^2\over \zeta}\right)\,\psi^{(1)}(\zeta)
		+\mathcal{K}^4 \psi (\zeta)=0\,,
	\end{equation}
	where $\mathcal K$ is a constant which depends upon the mode expansions along the directions $x^i$. 
	At the leading order in $\zeta$ (that is close to the AdS boundary) the radial differential equation \eqref{radialODE-ads4} becomes 
	\begin{equation}\label{radialODE-ads4-leading}
		\psi^{(4)}(\zeta)
		+\frac{2}{\zeta}\psi^{(3)}(\zeta)
		-\frac{1}{\zeta^2} \psi ^{(2)}(\zeta)
		+\frac{1}{\zeta^3}\,\psi^{(1)}(\zeta)=0\,.
	\end{equation}
	The main feature to notice here is the absence of a potential term at the leading order, in contrast with our case~\eqref{4th-kerr-leading-z0}. 
	Applying the Frobenius method, that is assuming $\psi(\zeta)=\zeta^\beta \sum_{n=0}^\infty c_n \zeta^n$, we obtain at the leading order the roots%
	\footnote{These roots are nothing but the roots of the equation $\Delta(\Delta-d)=m^2$, for a scalar field with mass $m$ in AdS$_{d+1}$. Here we have set $m=0$, and use $\beta$ instead of $\Delta$ to be consistent with the notation used in this work.}
	\begin{equation}\label{multiplicity}
		\beta=0, d\,, \qquad \text{with multiplicity} ~2\,. 
	\end{equation}
	The non-trivial multiplicity means that we obtain the following four linearly independent solutions
	\begin{equation}\label{AdSlogsol}
		\zeta^0,~~ \zeta^d, ~~\zeta^0 \log \zeta, ~~\zeta^d \log \zeta\,. 
	\end{equation}
	At the next-to-leading order in small $\zeta$, the differential equation becomes 
	\begin{equation}\label{radialODE-ads4-nexttol}
		\psi^{(4)}(\zeta)
		+\frac{2}{\zeta}\psi^{(3)}(\zeta)
		-2\left( \mathcal{K}^2+\frac{1}{\zeta^2}\right) \psi^{(2)} (\zeta)
		+\left(\frac{1}{\zeta^3}-{2\mathcal{K}^2\over \zeta}\right)\,\psi^{(1)}(\zeta)
		=0\,.
	\end{equation}
	The potential term is not present in the equation, but this is specific to the case $d=2$. 
	This equation can be solved analytically, and the four linearly independent solutions are Bessel functions of the first and second kind ($I_n, Y_n$), whose arguments depend on the dimensions of AdS and the constant $\mathcal K$, and logarithm. Again expanding these solutions around $\zeta=0$ the Bessel function $Y_n$ gives rise to another explicit logarithmic behavior close to the boundary, in agreement with the leading behaviour found in \eqref{AdSlogsol}. 
	
	With this in mind, we take inspiration from the action \eqref{action-laplacian-n} to start our investigation of higher derivative models in a Kerr black hole background and their hidden symmetries.
	
		\subsubsection{Kerr}

		We start this section by rewriting the radial Klein-Gordon equation in new coordinates, and then we examine the squared Klein-Gordon operator in this setting.
		
		Defining 
		\begin{equation}\label{z-coords}
			z=\frac{r-r_-}{r-r_+}   \,,
		\end{equation}
		the radial equation \eqref{KleinGordon} becomes
		\begin{equation}\label{KG-z}
			\begin{aligned}
				& z(1-z) f^{\prime\prime}(z)+(1-z)f^\prime(z)
				\\ 
				& -\left(\alpha_+^2-{\alpha_-^2\over z}+{K\over 1-z}-\left(\frac{(r_--r_+)^2}{(1-z)^3}+2{(r_--r_+)(M+r_+)\over (1-z)^2}+\frac{4 M^2+2 M r_++r_+^2}{1-z}\right)\omega^2\right)f(z)=0\,,
			\end{aligned}
		\end{equation}
		where we have used the definitions \eqref{alphas} in the above equation. 
		The singular points of the original equation \eqref{KleinGordon}, namely $r=r_-$, $r=r_+$, and $r=\infty$ have been mapped to $z=0$, $z=\infty$ and $z=1$ respectively. In these coordinates \eqref{z-coords} the Frobenius analysis becomes more transparent, and it is clear from equation \eqref{KG-z} that, for example at the leading order, close to the regular singular point $z=0$ ($r=r_-$) the equation is simply
		\begin{equation}
			f^{\prime\prime}(z)+\frac 1 z f^\prime(z)+\frac{\alpha_-^2}{z^2} f(z)=0\,.
		\end{equation}
		The two linear independent solutions are then 
		\begin{equation}
		z^{i\alpha_-}, \qquad z^{-i\alpha_-}\,. 
		\end{equation}
		
	Before moving to higher order differential equations, it is useful to examine the second order Klein-Gordon equation \eqref{KG-generics} for generic spin $s$ in this coordinate system. 
	\begin{equation}\label{KG-z-generals}
	    \begin{aligned}
	        	& z(1-z) f^{\prime\prime}(z)+(1-z)f^\prime(z)
				\\ 
				& -\left((\alpha^s_+)^2-{(\alpha^s_-)^2\over z}+{K+s^2\over 1-z}
			 +2 i s \left({ M-r_+ \over 1-z}-{r_+-r_-\over (1-z)^2}\right) \omega\right.
				\\
				&
				-\left.\left(\frac{(r_--r_+)^2}{(1-z)^3}+2{(r_--r_+)(M+r_+)\over (1-z)^2}+\frac{4 M^2+2 M r_++r_+^2}{1-z}\right)\omega^2\right)f(z)=0\,.
	    \end{aligned}
	\end{equation}
	Notice that the terms proportional to $\omega^2$ are unaffected by the spin $s$, while now there is a linear term proportional to $\omega$ and $s$. Again, expanding at leading order for example around $z=0$ ($r=r_-$), we obtain 
		\begin{equation}
			f^{\prime\prime}(z)+\frac 1 z f^\prime(z)+\frac{(\alpha^s_-)^2}{z^2} f(z)=0\,,
		\end{equation}
		where $\alpha_-^s$ is defined in equation \eqref{alphaspins}. The two independent solutions to this equation are
		\begin{equation}
		    z^{i\alpha_-^s}\,, \qquad z^{-i\alpha_-^s}\,. 
		\end{equation}
		Similarly, we can consider the next-to-leading order expansion of the full equation, and we have
			\begin{equation}\label{eq:nexttoleading}
			\begin{aligned}
			&f^{\prime\prime}(z)+\frac 1 z f^\prime(z)
			\\
			&+\left(\frac{(\alpha^s_-)^2}{z^2} -{K+s^2+(\alpha_+^s)^2-(\alpha_-^s)^2+2 i s (M-r_-)\omega-(4 M^2+2 M r_-+r_-^2)\omega^2\over z}\right)f(z)=0\,.
			\end{aligned}
		\end{equation}
		Notice that $$
		(\alpha_++\alpha_-)(\alpha_+-\alpha_--i s)=(\alpha_+^s)^2-(\alpha_-^s)^2\,. 
		$$
		The solutions to equation \eqref{eq:nexttoleading} are modified Bessel function of the first kind $I_n(z)$. 
		Continuing the expansion of equation \eqref{KG-z-generals} at the next-to-next-to-leading order, when constant terms appear in the potential, we see that the solutions are hypergeometric functions. We should stress that these equations are only valid in a neighborhood of $z=0$, and so the solutions obtained in this way are not the full solution of the original equation \eqref{KG-z-generals}. 
		
		Let us now consider equation \eqref{genreq} for the case $n=2$. Our starting point are equations \eqref{radialKG} and \eqref{radialKG-Phi} (after acting on the field $\Phi$). Applying again the Klein-Gordon operator \eqref{radialKG} to equation \eqref{radialKG-Phi}, and changing coordinate system as in \eqref{z-coords}, we obtain a rather lengthy expression. We then choose a constant $\theta$ slice, since as discussed in Section \ref{sec:Hidden-symm-HD} this does not affect the monodromy data. Again the singular points are $z=0$, $z=\infty$ and $z=1$, and by performing a series expansion around the regular singular points $z=0$ ($r=r_-$), $z=\infty$ ($r=r_+$) we obtain the roots \eqref{alphaminn2} and \eqref{alphapn2} respectively. 
		
		We find instructive to examine the fourth-order differential equation obtained in this way, close to a regular singular point at the leading order. We can focus on $z=0$ ($r=r_-$) for simplicity. Then the differential equation is given by 
		\begin{equation}\label{4th-kerr-leading-z0}
			\begin{aligned}
				f^{(4)}(z)+{4\over z} f^{(3)}(z)+\frac{2(1+\alpha_-^2)}{z^2} f^{(2)}(z)
				+\frac{\alpha_-^2(1+\alpha^2_-)}{z^4}f(z)=0\,,
			\end{aligned}
		\end{equation}
		and the four independent solutions are 
		\begin{equation}\label{highernkerralphas}
			z^{1-i\alpha_-}\,,~ z^{1+i\alpha_-}\,,~ z^{-i\alpha_-}\,,~ z^{i\alpha_-}\,.     
		\end{equation}
		At the next-to-leading order the structure of the equation is
		\begin{equation}\label{ntl-order4-z}
			\begin{aligned}
				&f^{(4)}(z)+\left({4\over z}+A_3(\theta_0)\right) f^{(3)}(z)
				+\left(\frac{2(1+\alpha_-^2)}{z^2}+\frac{A_2(\theta_0)}{z}\right) f^{(2)}(z)
				+\frac{A_1(\theta_0)}{z^2} f^{(1)}(z)
				\\ \noindent
				&+\left(\frac{\alpha_-^2(1+\alpha^2_-)}{z^4}
				+\frac{A_0(\theta_0)}{z^3}\right)f(z)=0\,.
			\end{aligned}
		\end{equation}
	 The constants $A_0, A_1, A_2, A_3$ depend on the black hole parameters, the dynamical inputs $K, \omega, m$, as well as the choice of the constant slice $\theta_0$, as our notation underlines. We refrain to write their explicit expression here, since it is not particularly useful. 
		The four linear independent solutions of equation \eqref{ntl-order4-z} are hypergeometric functions and Meijer G-functions, which again depend on the monodromy data and the constant $A$s. 
		
		We now need to compare and contrast the AdS case \eqref{multiplicity} and the Kerr case \eqref{highernkerralphas} in order to determine whether an analogous formulation of a Kerr/logCFT is possible. We know from Section \ref{sec:modelinads} that the fourth order radial equation in an AdS background admits logarithmic solutions \eqref{AdSlogsol}. In fact, the logarithmic behaviour is \textit{guaranteed} at the leading order close to the boundary by the degeneracy of the indicial roots, see \eqref{radialODE-ads4-leading}-\eqref{multiplicity}. However, the Kerr case is more subtle. The crucial difference between equations \eqref{radialODE-ads4-leading} and \eqref{4th-kerr-leading-z0} is the presence of a potential term in the Kerr black hole geometry.%
		\footnote{We should stress that we are only discussing about formal similarities. In the Kerr black hole case we zoom in a region close to the horizon, while in the AdS case, we are interested in a boundary behaviour, where the CFT lives.}
		We can see from equation \eqref{highernkerralphas} that two pairs of indicial roots differ by a positive integer. This signals that a logarithmic solution \textit{may or may not} be present. In the Kerr/CFT construction, the CFT exists at the black hole horizon, and so to determine whether the logarithmic terms are there in the region that interests us, we expand the Meijer G-functions close to $z=0$. 
		A general expansion of this function contains terms polynomial in $z$ and also terms like $z^2 \log z$. The corresponding coefficients are very lengthy expressions, which depend on the $A$ constants.
		We remind the reader that these are not solutions of the full equation \eqref{KG-z}, their validity is within the validity of the expansion of the equation \eqref{ntl-order4-z} itself, and so if the solution shows a logarithmic term beyond its perturbative regime is not meaningful in this context. We have investigated the coefficient of the logarithmic term numerically, and interestingly, we were not able to find a non-zero coefficient near the black hole horizon.
		This seems to indicate that a Kerr/logCFT construction is \textit{not} possible within this framework. We discuss physical interpretations of this result in the discussion section. 
	It is perhaps worth stressing that a logarithmic solution can still be present outside of our regime of validity, that is at order $z^2$. Again, the presence or not of a logarithmic term in general depends on the specific dynamics, in particular on the slice $\theta_0$ which we chose. We remind the reader that the full fourth-order equation \eqref{KG-z} is not separable, and we choose a specific $\theta$-slice to perform our analysis. This is in contrast to the monodromy data, which do not depend on this choice.


	\section{Discussion}\label{sec:discussion}
	
 The goals of this article were to provide a case study of how hidden conformal symmetry is manifest when we change the dynamics on a given background, and in particular whether we could use variations of the Kerr/CFT correspondence to work toward diagnosing a new instance of a logCFT correspondence. In this section we review our results, discuss the challenges with a logCFT construction, and discuss future directions. 
	
	We found the monodromy parameters $\alpha^{(n)}$ for general number of derivatives $2n$, and show how they are related to the monodromy exponents of the regular Klein-Gordon equation of higher spin fields. We show that pairs of the indicial roots $i\alpha^{(n)}$ differ by an integer, and thus a logarithmic contribution to the radial equation could be present. However, we find that sufficiently close to the black hole horizon, potential logarithmic contributions vanish. This seems to indicate that we cannot construct a Kerr/logCFT correspondence from higher derivative theories, which have been used to construct examples of AdS/logCFT correspondences. 
	
	The difficulty in constructing a Kerr/logCFT correspondence is interesting, both from the gravitational perspective and the field theory perspective. From the field theory perspective, logCFTs have proven to be relevant in numerous areas of physics. Indeed, they can arise at critical points of various physical systems, such as those describing quantum Hall plateau transition~\cite{Flohr:1995zj, Gurarie:1997dw, Cappelli:1998ma, Ino:1998xe}, but also in models describing percolation~\cite{Saleur:1991hk}, self-avoiding walks~\cite{Duplantier:1987sh}, and systems with quenched disorder~\cite{Cardy:1999zp, Caux:1995nm, Maassarani:1996jn}. 
	These special conformal field theories are characterised by a logarithmic behaviour in correlation functions~\cite{Saleur:1991hk, Rozansky:1992td, Gurarie:1993xq}, which seems to clash with the fact that the theory is scale invariant. However, the presence of these terms is hinged on the reducible but indecomposable representations of the conformal group~\cite{Gurarie:1993xq}. The crucial point is that the conformal Hamiltonian is not diagonalizable, but rather has a Jordan cell structure (for rank $n\ge 2$), which leads to logarithmic terms in the correlation functions, and to lack of unitarity. 
	While this feature would be generally considered a red flag in quantum field theory, it does not pose any threat as a description of statistical mechanical systems, as confirmed by the examples mentioned above. We refer the reader to {\it e.g.}~\cite{Hogervorst:2016itc, Grumiller:2013at, Cardy:2013gk}, and references therein, for more recent and extensive reviews on logCFTs. 
	
	From a gravitational perspective, a natural question is whether hidden conformal symmetries are still visible (or modified) when the dynamics is encoded in higher derivative differential operators, as those in the action \eqref{action-laplacian-n}. Higher derivative theories breaks unitarity, hence we might expect that this would be reflected somehow in the hidden symmetry group. Consequently, this might  hinder us in our efforts to investigate/study a Cardy-like formula in this setting. Indeed, if we expect a non-unitary CFT the partition function might not be bounded from below, we might have states with negative norm, thus it is not clear in which sense we could discuss of an entropy. Still, it might be possible to give a description of the density of states, perhaps taking into account anomalies. 
	
	There are still further challenges with trying to make a logCFT correspondence in the spirit of Kerr/CFT using the model we propose. First, there is still some work to be done regarding making a robust holographic correspondence in Kerr/CFT itself. Even though there is ample evidence that the hidden conformal symmetry found in \cite{Castro:2010fd} really is described by an underlying CFT (such as the correct computations of scattering cross-sections), many elements are still lacking, such as how to conduct an asymptotic symmetry group analysis when the symmetry generators are not all isometries, and when the conformal symmetry acts at the horizon and not the boundary. We leave this interesting problem for future work. Also, as mentioned previously, in the case of a logCFT correspondence in AdS the scalar field is guaranteed to have a logarithmic piece near the boundary, but in Kerr the logarithmic piece might vanish near the horizon. 
	
	 Lastly, we illustrate once again that the monodromy method is really a powerful tool for studying hidden conformal symmetry: it allows us to study near horizon dynamics without actually taking a near-region limit in the dynamics (although sometimes we do take such a limit, purely for calculational ease). Furthermore, even though the higher order equations $\left(\nabla^\mu\nabla_\mu\right)^2\Phi=0$ are no longer separable, it does not matter for the monodromy analysis. Our results are independent of the $\theta$ slice we choose.

There are several interesting opportunities for future work. One important contribution would be to establish a non-extremal analog to the asymptotic symmetry group analysis presented for example in \cite{Guica:2008mu}. This would further strengthen the claim of a non-extremal Kerr/CFT correspondence. Another direction is that this paper is a first step towards learning what the monodromy method can tell us other equations of motion. It would be interesting, for example, to study whether hidden conformal symmetry is somehow encoded in the Dirac or geodesic equations. Further, it was recently shown in \cite{Keeler:2021tqy} that there is a difficulty in constructing conformal coordinates in six spacetime dimensions and higher that do not have branch cuts, unless an explicit near-horizon limit is taken. Thus it would be interesting to check whether the general form on the Klein-Gordon operator discussed in Section \ref{sec:standardform} and Appendix \ref{app:KGgoodform} holds in higher dimensions. We leave this for future work.

An intriguing and open question is whether a generalisation of the Cardy-formula exists for non-unitary theories, particularly those where the underlying conformal field theory is a logarithmic one ~\cite{Grumiller:2013at}. Indeed, exploring this question was one of our initial motivations for this work. 	The Cardy formula has played a crucial role in the AdS/CFT duality~\cite{Maldacena:1997re, Witten:1998qj, Gubser:1998bc}, and in particular in the Kerr/CFT correspondence~\cite{Guica:2008mu}.  The  Bekenstein-Hawking entropy of asymptotically AdS$_3$ black hole exactly reproduces (at high energy) the degeneracy of states governed by the Cardy formula~\cite{Cardy:1986ie} in two-dimensional CFTs, that is
	\begin{equation}\label{S-cardy}
		S_{CFT}= 2\pi \sqrt{{c_R L_0\over 6}}+2\pi \sqrt{{c_L \bar L_0\over 6}}\,,
	\end{equation}
	where $c_R, c_L$ are the right and left central charges, respectively, and $L_0, \bar L_0$ are the zero-th generators of the Virasoro algebra. 
	It was later understood that  any higher-dimensional black holes with an AdS$_3$ near-horizon geometry will obey a Cardy-formula, again as a consequence of the symmetry of the given geometry. 
	A step further was made in \cite{Detournay:2012fk} where the Cardy formula was extended to warped AdS$_3$ geometry, that is geometries with $\grSL(2,\R)\times \grU(1)$ isometries. 
	In particular, these are the isometries of the near-horizon geometry of the extremal Kerr black holes (and in general of extremal/near extremal black holes). 
	Two essential ingredients enter in the derivation of the Cardy formula \eqref{S-cardy}: unitarity and modular invariance, and so the presence of a Cardy formula, even with a reduced symmetry group as in the warped  case, tells us that there is still a notion of modular invariance here.  We leave the establishment of a Cardy-like formula in non-unitary settings for future work.

	\section*{Acknowledgements}
	
	We are indebted to Cynthia Keeler and Rahul Poddar. 
	We thank Prof. Norma Sanchez for inviting us to contribute with this article to the Open Access Special Issue “Women Physicists in Astrophysics, Cosmology and Particle Physics”, published in [Universe] (ISSN 2218-1997) and to M. Grana, Y. Lozano, S. Penati and M. Taylor for involving us in this special issue. 
	This research was supported in part by the Icelandic Research Fund under contract 195970-052 and by grants from the University of Iceland Research Fund.

	\begin{appendix}
		
		\section{Other standard form examples}\label{app:KGgoodform}
		
		In this short appendix we would like to discuss a general form for the d'Alembertian operator $\box\equiv\nabla_\mu \nabla^\mu$ that persists when acting on fields of higher spin and in higher dimensions. We present this discussion because this form was useful to us in examining monodromies in theories with higher-order equations of motion $(\nabla_\mu\nabla^\mu)^n\Phi=0$, and because we feel that it highlights important physical structure related to monodromy parameters and conformal coordinates defined in \eqref{seriessol} and \eqref{confcoordgen} respectively.
		
		To set the stage, let's state the results for the four-dimensional case: a scalar field $\Phi=e^{i(m\phi-\omega t)}R(r)S(\theta)$ propagating on a Kerr background \eqref{kerrMet}. The Klein-Gordon equation is  
		\begin{equation}\label{appKG}
			\nabla^{\mu}\nabla_{\mu}\Phi=\frac{\Phi}{\rho^2R(r)}	\left[\partial_r(\Delta\partial_r)+\alpha_+^2\frac{g^2(r)}{f^2(r)}-\alpha_-^2g^2(r)+(r^2+2Mr)\omega^2+C_{t,\phi}\right]R(r),
		\end{equation} 
		where, as we discussed in Section \ref{sec:Hidden-symm-KG},
		\begin{equation}\label{4dradial}
			\alpha_{\pm}=\frac{\omega-\Omega_\pm m}{2\kappa_\pm}
		\end{equation}
		are the monodromy parameters, the functions
		\begin{equation}
			f(r)=\left(\frac{r-r_+}{r-r_-}\right)^{1/2}, \qquad g(r)=\left(\frac{r_+-r_-}{r-r_-}\right)^{1/2}
		\end{equation}
		define the radial dependence of the conformal coordinates, and $C_{t\phi}$ is a constant of motion, see equations \eqref{CarterOp}-\eqref{radialKG} and discussion below. In scenarios with higher spin fields and higher-dimensional spacetime backgrounds, the quantities $(\alpha_\pm, f(r), g(r), C_{t,\phi})$ change, but the overall form of \eqref{appKG} does not. Let's see how this works.
		
		The hidden conformal symmetry generators for five-dimensional Myers-Perry black holes \cite{myers1986black} were first studied by \cite{Krishnan:2010pv}. The radial equation of motion for a scalar field ansatz $\Phi=e^{i(-\omega t+m_1\phi_1+m_2\phi_2)}$ can be written as
		\begin{equation}\label{KG-generics}
			\left[\frac{\partial}{\partial x}\left(x^2-\frac{1}{4}\right)\frac{\partial}{\partial x}+\alpha_+^2\frac{g^2(r)}{f^2(r)}-\alpha_-^2g^2(r)+\frac{x\Delta\omega^2}{4}+\tilde{C}_{t,\phi}\right]\Phi=0,
		\end{equation}
		where 
		$$x\equiv\frac{r^2-1/2(r_+^2+r_-^2)}{(r_+^2-r_-^2)}$$ is a radial coordinate and the monodromy parameters $\alpha_\pm$ were found in Appendix A of \cite{Castro:2013kea}. There are two important points here. First, the functions $f(r)$ and $g(r)$ are precisely those that define the radial behavior of the conformal coordinates in a five-dimensional setting, presented in \cite{Krishnan:2010pv}. Second, we can see a pattern emerging. The Klein-Gordon equation is expressible as a derivative piece, pieces involving the monodromy parameters (the form of which are fixed), a constant term and a non-constant $r$-dependent term that is irrelevant neither either horizon. 
		
		The monodromy analysis for higher spin $s$ perturbations on a four-dimensional Kerr background was treated in \cite{Castro:2013lba}. The equation of motion for such a perturbation can be written as
		\begin{equation}
			\left(\partial_r\Delta\partial_r+(\alpha^s_+)^2\frac{g^2(r)}{f^2(r)}-(\alpha^s_-)^2g^2(r)+\omega^2r^2+2(M\omega+is)\omega r+\mathcal{C}_{t,\phi}\right)R(r)=0.
		\end{equation}
		As expected, the functions $f(r)$ and $g(r)$ attached to the monodromy parameters $\alpha_\pm$ are the same as in \eqref{4dradial}, and the monodromy parameters themselves are\footnote{There is a typo in the $\alpha_\pm$ reported in \cite{Castro:2013lba}.}
		\begin{equation}\label{alphaspins}
			\alpha^s_{\pm}=\mp\frac{is}{2}+\frac{2M\omega r_{\pm}-am}{r_+-r_-}.
		\end{equation}
		Notice that setting $s=2$ in the expressions \eqref{alphaspins}, the monodromy parameters reduce to \eqref{alphapn2}-\eqref{alphaminn2}.

	\end{appendix}

	\bibliographystyle{utphys2}
	\bibliography{ConformalFromKilling}

\end{document}